\documentclass[prd,twocolumn,nofootinbib]{revtex4-1}
\pdfoutput=1
\usepackage{amsmath,amssymb,graphicx}
\usepackage{epsfig}

\begin{document}

\preprint{\hepth{0910.1397}}

\title{Stretched extra dimensions and bubbles of nothing in a toy model 
landscape.}

\author{I-Sheng Yang}
\email[E-mail me at: ]{isheng.yang@gmail.com}
\affiliation{ISCAP and Physics Department \\
Columbia University, New York, NY, 10027 , U.S.A.}

\begin{abstract}
Using simple 6D junction conditions, we describe two surprising 
geometries.  First in a case of transitions between $dS_4\times S_2$ vacua, 
the $S_2$ can be stretched significantly larger than the vacuum values both 
before and after the transition.  Then we discover that the na\"{i}ve 
instability to decompactification is actually a bubble of nothing instead. 
\end{abstract}

\maketitle

\section{Introduction and outline}

The six dimensional Einstein-Maxwell theory\cite{FreRub80,RanSal82} provides
stable compactifications to $M_4\times S_2$, where $M_4$ can be deSitter, 
anti-deSitter or Minkowski space.  Recently it is recognized as a good toy 
model to study transitions between different vacua with compactified 
extra dimensions\cite{DouKac06,BlaJos09,CarJoh09}, a scenario that arises 
from the landscape of string theory\cite{BP}.

One advantage of this model is that we can have a lot of 4D vacuum solutions 
with similar sizes of $S_2$.  Therefore, it is natural to assume that $S_2$ 
remains in a similar size during vacuum transitions, as we usually do in more 
detailed models.  However, Johnson and Larfors in\cite{JohLar08} pointed out 
a problem of such assumption in a string theory model.  

In Sec.\ref{sec-problem}, we will give an intuitive picture explaining why
freezing the extra dimensions during vacuum transitions is generally not a 
good idea.

In Sec.\ref{sec-GR}, we go over general equations in the 6D Einstein-Maxwell 
theory.  In particular how to describe the vacuum solutions and transitions 
between them from a purely geometric point of view.

In Sec.\ref{sec-4t4}, we study 4D to 4D transitions with the geometric method 
and the conventional dimensional reduced method side by side.  Together they
allow us to describe how the extra dimension changes during the transition.  
Choosing some allowed values of a free parameter ensures that the extra 
dimensions are stretched during the process.  

In Sec.\ref{sec-4t6}, we use the same construction for a new 4D to 6D 
transition.  Different from both the compactification\cite{CarJoh09} and
the decompactification\cite{GidMye04}, our solution does not have 6D 
asymptotics.  It is similar to a bubble of nothing\cite{Wit81} within a 4D 
vacuum.

Finally, in Sec.\ref{sec-dis} we summarize and comment on possible future 
directions.  The stretched extra dimensions may change our picture on bubble
collisions\cite{EasGib09,BlaJos09}, and help to clarify whether in some cases
the transition will be forbidden\cite{JohLar08}.  The bubble of nothing
\cite{Wit81} may replace the decompactification to be the universal 
instability in models with extra dimensions.

\section{Better not to freeze the extra dimensions}
\label{sec-problem}

\begin{figure}
\begin{center}
\includegraphics[scale = 0.6, angle=0]{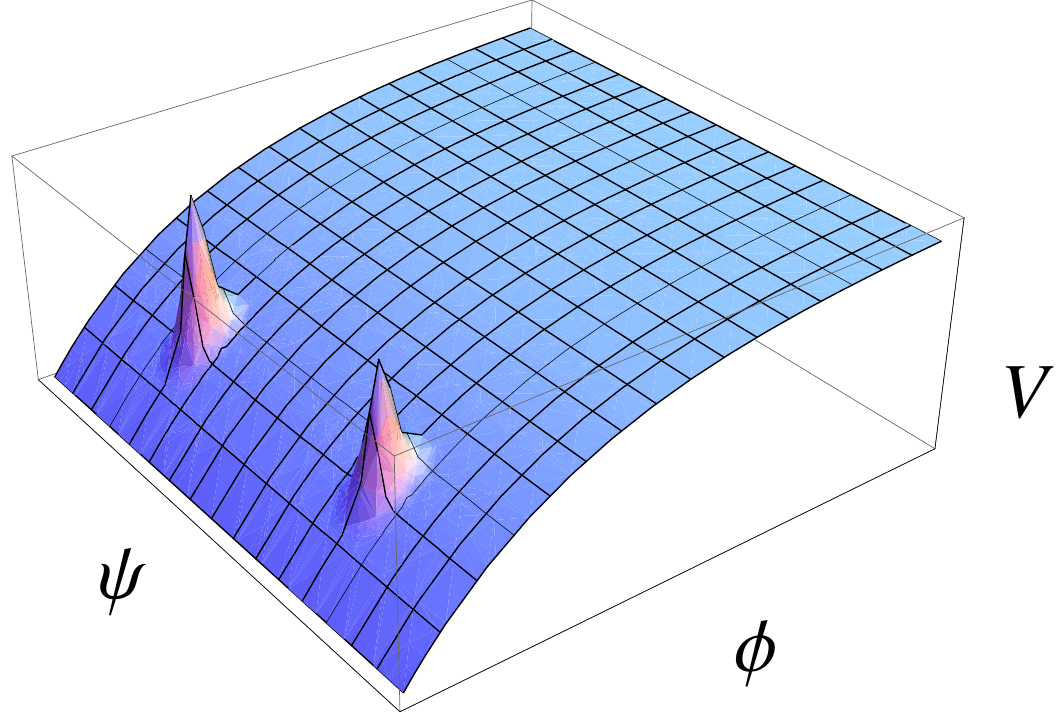}
\end{center}
\caption{An effective Euclidean potential for a theory with extra dimensions
and discrete vacua.  $\phi$ is related to the size of extra dimension, so 
there is a decompactification direction where $V$ is asymptotically zero.  In
2 different values of $\psi$ but very similar values of $\phi$ we have two 
local maxima---the vacua.} 
\label{fig-needles}
\end{figure}

Fig.\ref{fig-needles} is a typical effective (Euclidean) potential of a 
theory with extra dimensions and discrete vacua.  It contains the following 2 
traits.
\begin{itemize}
 \item In the $\phi$ direction, the potential gradually slopes to zero, which 
       corresponds to decompactifying the extra dimensions.  This can be the 
       real part of K\"{a}hler moduli for string theory\cite{KKLT}.\\
 \item Discrete vacua distribute not only in the above direction, but also
       in some other directions $\psi$.  For example the imaginary part of 
       K\"{a}hler moduli in\cite{BlaBur04}.  \\
\end{itemize}
Because the discrete vacua are stablized by non-perturbative effects, for the
value $\psi$ not supporting any vacuum, the effective potential typically 
follows the general slope in $\phi$.  

Now consider 2 vacua with the same value of $\phi$, namely the same size of 
extra dimenions.  Fixing that size during a transition implies connecting 
the 2 vacua by a straight line in Fig.\ref{fig-needles}.  This line certainly
does not obey any equation of motion, as it should be swept away by the 
non-zero slope in the $\phi$ direction.  A more reasonable path representing
the transition would be like a projectile motion---climb the slope in 
$\phi$ direction and come back, in the meanwhile move from one vacuum to 
another.  This implies the extra dimension during a transition can be very 
different from its vacuum value.

Such transition might be very annoying to model, since a generic 
multidimensional effective potential will not be as simple as 
Fig.\ref{fig-needles} and one can only search for the path numerically.  
In the following 2 sections we will show that in the 6D Einstein-Maxwell 
theory, there is a simple way to monitor transitions with significant changes 
in the extra dimensions.

\section{6D spacetime with vacuum energy and flux}
\label{sec-GR}
 \subsection{Stable solutions}
 \label{sec-soln}
We consider the 6D Einstein-Maxwell theory with positive vacuum energy 
$\Lambda_6$ and 4-form fields, as it is necessary to ensure $dS_4\times S_2$ 
compactifications. For the solutions we will use in this paper, it is most 
convenient to write down a general metric with $SO(3,1)\times SO(3)$ symmetry.
\begin{equation}
ds^2=d\rho^2+A^2(\rho)dS_3^2+B^2(\rho)d\Omega_2^2~.
\end{equation}

\begin{figure}
\begin{center}
\includegraphics[scale = 0.4, angle=-90]{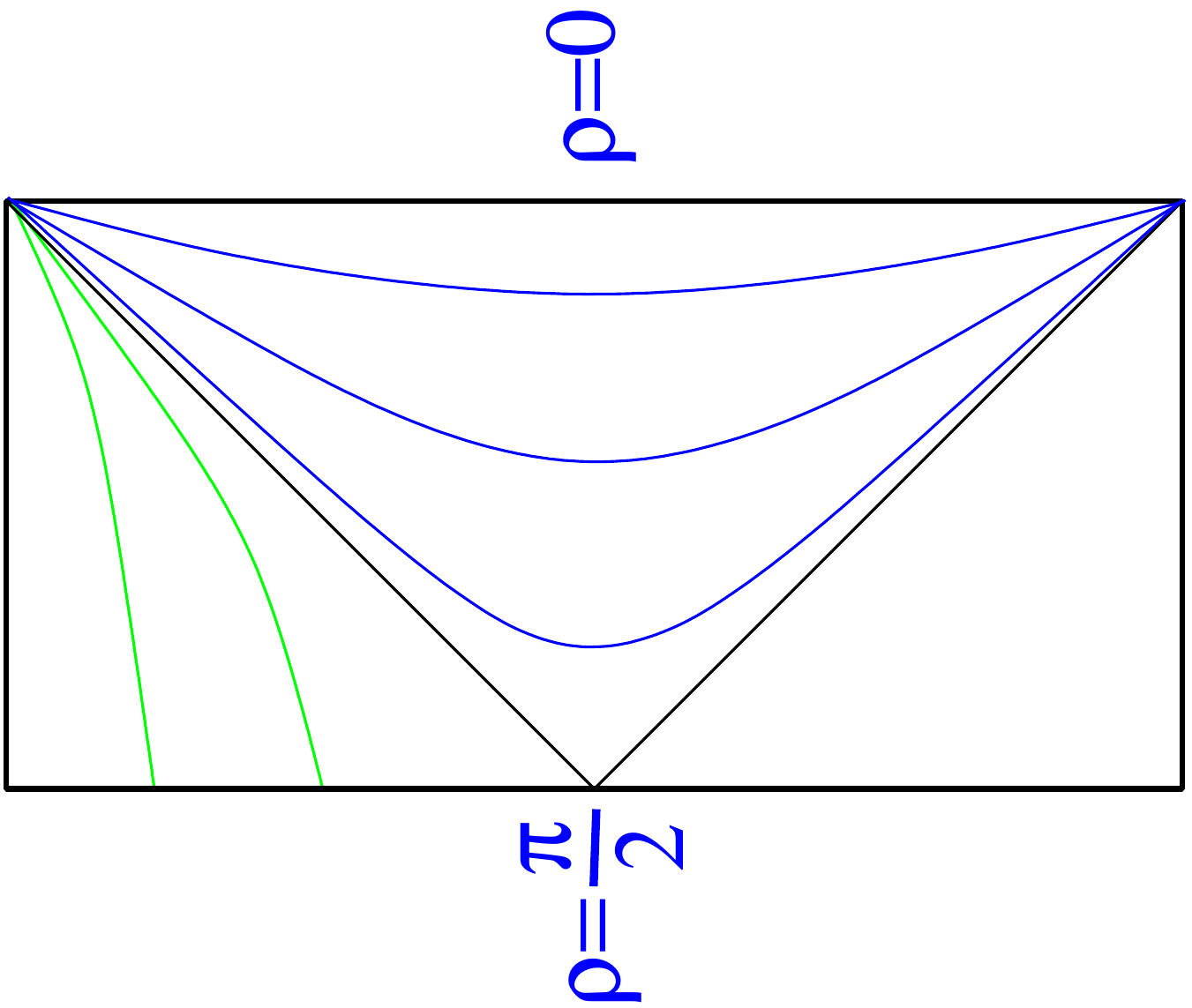}
\end{center}
\caption{The special Penrose diagram for $dS_6$ where two $S_2$s are
suppressed individually.  One of them has zero size on the left boundary, and
the other one is zero on the right boundary.  Our coordinate covers the
region with blue (timelike) slices where the physical radius of one $S_2$ ,
$B(\rho)$, is constant, and the radius of $dS_3$ formed by the other $S_2$
with time, $A(\rho)$, is also constant.  These are the slicings convenient
to put the charged brane.  The green slices are the similar constant size
surfaces with $B>L$, which can be obtained by analytically continuing our
coordinate.}
 \label{fig-6D}
\end{figure}

The Einstein equations are
\begin{eqnarray}
\frac{1}{m_6^4}\bigg(\Lambda_6 + \frac{Q^2}{4B^4}\bigg) &=& 3\frac{1-\dot{A}^2}{A^2}-6\frac{\dot{A}\dot{B}}{AB}+\frac{1-\dot{B}^2}{B^2}~, 
\label{eq-eeq1} \\
\frac{1}{m_6^4}\bigg(\Lambda_6 + \frac{Q^2}{4B^4}\bigg) &=&
\frac{1-\dot{A}^2}{A^2}-2\frac{\ddot{A}}{A}-4\frac{\dot{A}\dot{B}}{AB}
+\frac{1-\dot{B}^2}{B^2}-2\frac{\ddot{B}}{B}~, 
  \nonumber \\
\frac{1}{m_6^4}\bigg(\Lambda_6 - \frac{Q^2}{4B^4}\bigg)&=& 
3\frac{1-\dot{A}^2}{A^2}-3\frac{\ddot{A}}{A}-3\frac{\dot{A}\dot{B}}{AB}
-\frac{\ddot{B}}{B}~. \nonumber
\end{eqnarray}
Here $m_6$ is the 6D planck mass and $Q$ is the quantized charge as the 
field sources, namely 2-branes.  The 6D dS space is a solution with $Q=0$, 
which corresponds to 
\begin{eqnarray}
A(\rho)&=&L\cos\bigg(\frac{\rho}{L}\bigg)~,  \\
B(\rho)&=&L\sin\bigg(\frac{\rho}{L}\bigg)~, \label{eq-dS6}
\end{eqnarray}
where $0<\rho<\pi/2$, $10m_6^4L^{-2}=\Lambda_6$.  Of course we can 
interchange $A$ and $B$, it is still the same solution.

Nonzero $Q$ induces ``compactified'' solutions as 
\begin{eqnarray}
A(\rho)&=&\frac{\sin(H\rho)}{H}~, \\
B(\rho)&=&R~.
\end{eqnarray}
Here it is given in the convenient form for dS spaces, but we can easily take 
$H\rightarrow0$ for Minkowski space and imaginery $H$ for AdS spaces.  Using 
the Einstein equations, we can relate the size of the compactified dimensions 
$R$ and the 4D hubble constant $H$ to both $Q$ and 
$\Lambda_6$\footnote{Some of the solutions are 
unstable\cite{BouDew02,KriPab05}, but the solutions we use in 
Sec.\ref{sec-4t4} are all stable ones.}.
\begin{eqnarray}
3H^2+\frac{1}{R^2} &=&
 \frac{1}{m_6^4}\bigg(\Lambda_6+\frac{Q^2}{4R^4}\bigg)~,\\
6H^2 &=& \frac{1}{m_6^4}\bigg(\Lambda_6-\frac{Q^2}{4R^4}\bigg)~.
\end{eqnarray}

 \subsection{Transitions between different solutions}
 \label{sec-trans}
From the geometric point of view, vacuum transitions are related to geometries with solutions of different $Q$s patched together.  For example, 
two different solutions seperated by a charged 2-brane at 
($\bar{A},\bar{B}$).  Note that a boundary in 6D is a 5D object, but 2-branes 
are only 3D objects.  We have to sprinkle the branes in 2 of the dimensions 
like dust to construct a 5D boundary.

Such patched solutions must obey Israel junction conditions\cite{Isr66} at 
the boundary.
\begin{eqnarray}
2\bigg(\frac{\dot{A}_1+\dot{A}_2}{\bar{A}}\bigg)
+2\bigg(\frac{\dot{B}_1+\dot{B}_2}{\bar{B}}\bigg)&=&
\frac{1}{m_6^4}\bigg(\frac{\sigma}{4\pi\bar{B}^2}\bigg)~, \\
3\bigg(\frac{\dot{A}_1+\dot{A}_2}{\bar{A}}\bigg)
+\bigg(\frac{\dot{B}_1+\dot{B}_2}{\bar{B}}\bigg)&=& 0~.
\label{eq-juncq}
\end{eqnarray}
It is basically the integrated Einstein equations with a delta-function 
source.  $\sigma$ is the total tension (energy density in 2D) of the 2-branes
we sprinkled on the $4\pi\bar{B}^2$ sphere.  Here the convention is that in 
spacetime region $i$, the solution is ($A_i(\rho_i)$, $B_i(\rho_i)$) with the 
small $\rho_i$ region. ( namely, if $\dot{A_i}>0$, it means $A_i$ is 
increasing while moving toward the boundary. )  

Note the positivity constraint on the tension,
\begin{equation}
\frac{\sigma}{4\pi\bar{B}^2m_6^4 } = -4\bigg(\frac{\dot{A}_1+\dot{A}_2}{\bar{A}}\bigg)>0~.
\end{equation}

Combining Eq.~(\ref{eq-juncq}) with Eq.~(\ref{eq-eeq1}), we have
\begin{equation}
6\frac{\dot{A}_1^2-\dot{A}_2^2}{\bar{A}^2}
=\frac{Q_1^2-Q_2^2}{4m_6^4\bar{B}^4}~.
\label{eq-HQ}
\end{equation}

These tell us an important message.  For example, when $Q_1<Q_2$, we have
$\dot{A}_1^2<\dot{A}_2^2$.  So $\dot{A}_2$ needs to be negative to make the
tension positive.  More generally, from the side with the larger $Q$, this
boundary always looks like a small bubble (because the bubble radius is
shrinking while we approach the boundary from this side).

\section{4D to 4D vacuum transitions}
\label{sec-4t4}
 \subsection{Geometric point of view}
 \label{sec-4t4geo}
If we limit ourselves to ideal 4D vacua, where $H^{-1}\gg R$, we have
\begin{eqnarray}
\frac{m_6^4}{R^2} 
&=& 2\Lambda_6\bigg(1-\frac{9H^2m_6^4}{2\Lambda_6}\bigg)~,
\label{eq-R} \\
\frac{Q^2}{m_6^8} &=& 
\frac{1}{\Lambda_6}\bigg(1+\frac{3H^2m_6^4}{\Lambda_6}\bigg)~.
\label{eq-Q}
\end{eqnarray}

Note that for these vacua, $R\sim(m_6^4/2\Lambda_6)^{1/2}$ cannot change a 
lot.  This gives us a few advantages.  First of all, larger $H$ in our 6D 
point of view really means a larger effective 4-dimensional cosmological 
constant.  Eq.~(\ref{eq-HQ}) tells us larger $Q$ should be outside, which 
agrees with the usual tunneling picture that bigger cosmological constant 
should be outside.  There will be geometries with a clean seperation of 
scales, 
\begin{equation}
R^{-1}\gg\bar{A}^{-1}\gg H_2>H_1~,
\label{eq-scales}
\end{equation}
which represent standard thin-wall, small bubbles\cite{CDL} that does not 
care for the extra dimensions.  Within these solutions, we can simplify 
further calculations with
\begin{equation}
\dot{A}_1=-\dot{A}_2=1+O(H_i^2\bar{A}^2)~.
\label{eq-ass}
\end{equation}

Combine the junction condition Eq.~(\ref{eq-juncq}) and Eq.~(\ref{eq-HQ}), we 
have
\begin{equation}
\frac{\sigma}{4\pi\bar{B}^2m_6^4} =
\frac{m_6^8\bar{A}(H_2^2-H_1^2)}{4\Lambda_6^2\bar{B}^4}~.
\label{eq-tension}
\end{equation}

Furthermore, if $B$ does not change a lot during the transition, we have
$\bar{B}\sim R_i\sim R=(m_6^4/2\Lambda_6)^{1/2}$.  The entire transition can 
be understood from the 4D point of view.  Using the standard dimensional 
reduction method, we rescale length and planck mass as following,
\begin{eqnarray}
l_{4D} &=& l_{6D}Rm_6~, \nonumber \\
m_4 &=& \sqrt{4\pi} m_6~. \nonumber  
\end{eqnarray}
Eq.~(\ref{eq-tension}) reproduces the na\"ive junction condition in 4D
\begin{equation}
\sigma_{4D}=\frac{r_c\Delta\Lambda_{4D}}{3}
\end{equation}
as expected.

 \subsection{Effective 4D theory}
 \label{sec-4t4eff}

From the very beginning, we could have followed\cite{BlaJos09} and use the 
standard dimensional reduced effective 4D theory.
\begin{eqnarray}
\sqrt{4\pi}m_6    &=& m_4~, \nonumber \\
m_6 B(\rho)       &=& e^{\frac{\phi(\tau)}{2m_4}}~, \nonumber \\
m_6A(\rho)B(\rho) &=& a(\tau)~, \nonumber \\
m_6B(\rho)d\rho   &=& d\tau~. \nonumber
\end{eqnarray}

This translates the 6D Einstein equations into the general 4D FRW equations
and an equation of motion for the field $\phi$.
\begin{eqnarray}
\phi''+3\frac{a'}{a}\phi'  &=& \frac{\partial V}{\partial \phi}~,
\label{eq-field} \\
\bigg(\frac{a'}{a}\bigg)^2 &=&
\frac{1}{3m_4^2}\bigg(\frac{\phi'^2}{2}-V\bigg)+\frac{1}{a^2}~, \\
\frac{a''}{a}              &=& -\frac{1}{3m_4^2}\bigg(\phi'^2+V\bigg)~.
\end{eqnarray}
Here $'$ means the derivative to $\tau$ and the effective potential
\begin{equation}
V(\phi)=m_4^2\bigg(\frac{\Lambda_6}{m_6^4}e^{-\frac{\phi}{m_4}}
-m_6^2e^{-2\frac{\phi}{m_4}}+\frac{Q^2}{4}e^{-3\frac{\phi}{m_4}}\bigg)~.
\end{equation}

\begin{figure}
\begin{center}
\includegraphics[scale = 0.55, angle=0]{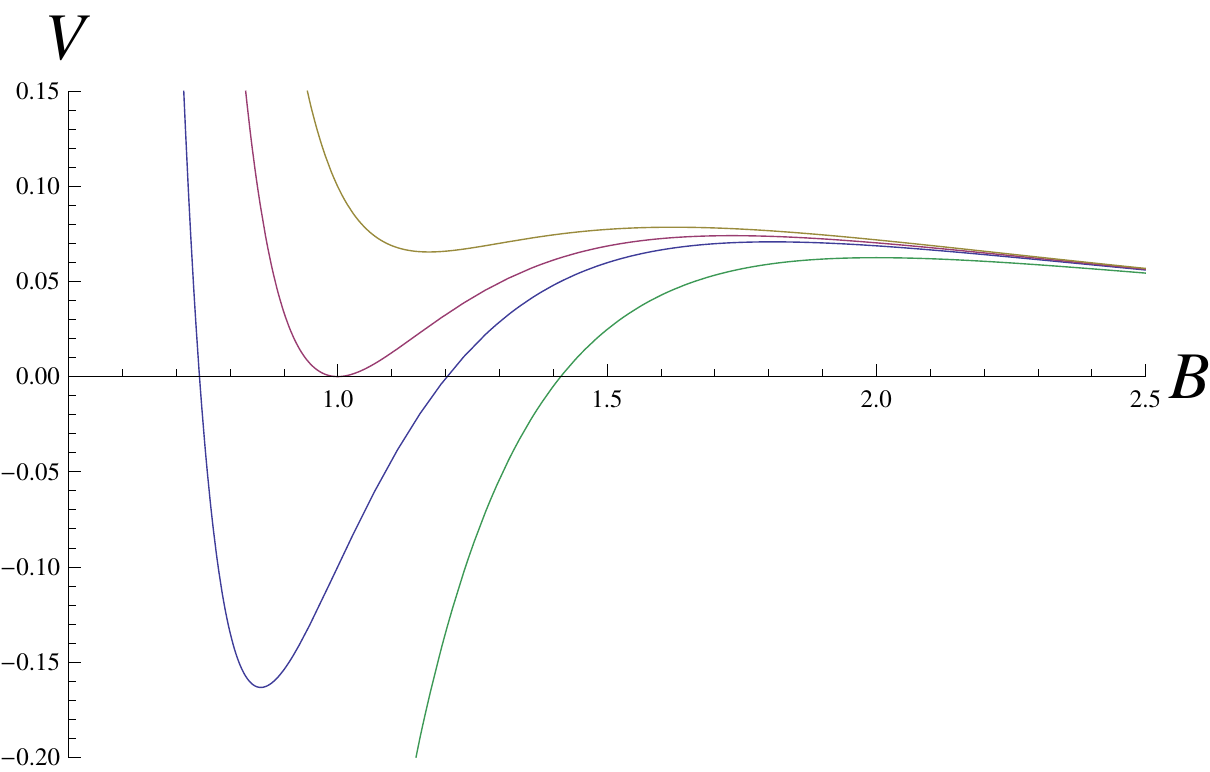}
\end{center}
\caption{
Effective potential as a function of $B=m_6^{-1}e^{\frac{\phi}{2m_4}}$.  
From high to low we have potentials with decreasing $Q$, the stablized 4D 
vacuum being deSitter, Minkowski and AdS.  The lowest one with $Q=0$ has no 
stablized 4D vacuum.}
 \label{fig-pot}
\end{figure}

\begin{figure}
\begin{center}
\includegraphics[scale = 0.35, angle=0]{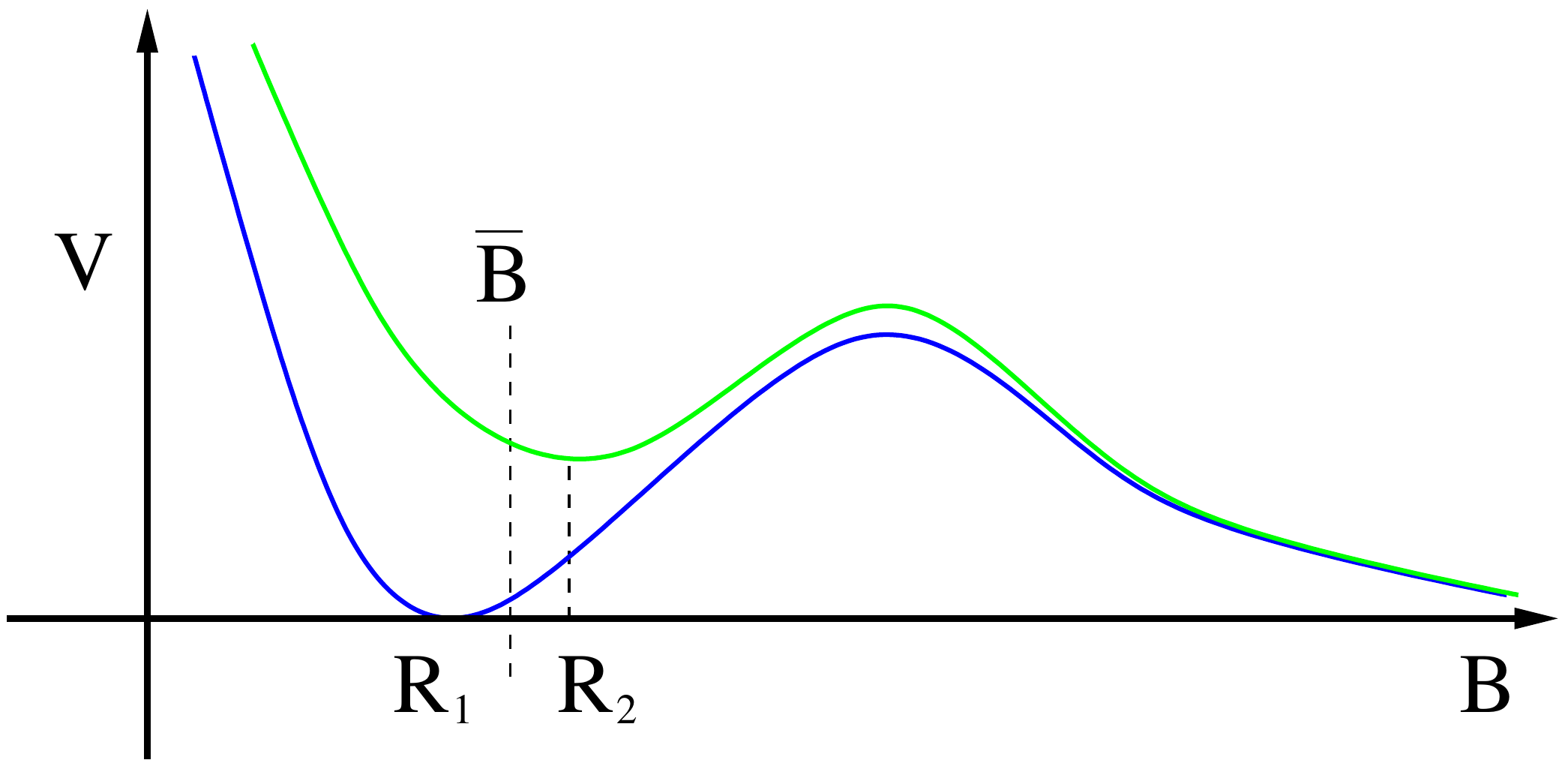}
\end{center}
\caption{The effective potential of two solutions with different charges, 
$Q_1$(blue, lower) and $Q_2$(green, higher).  Our solution corresponds 
to the radius $B$ started at $R_2$, moved through the green(higher) potential 
to $\bar{B}$ and hit a charged brane, then followed the blue(lower) potential 
to $R_1$.}
\label{fig-4t4}
\begin{center}
\includegraphics[scale = 0.27, angle=-90]{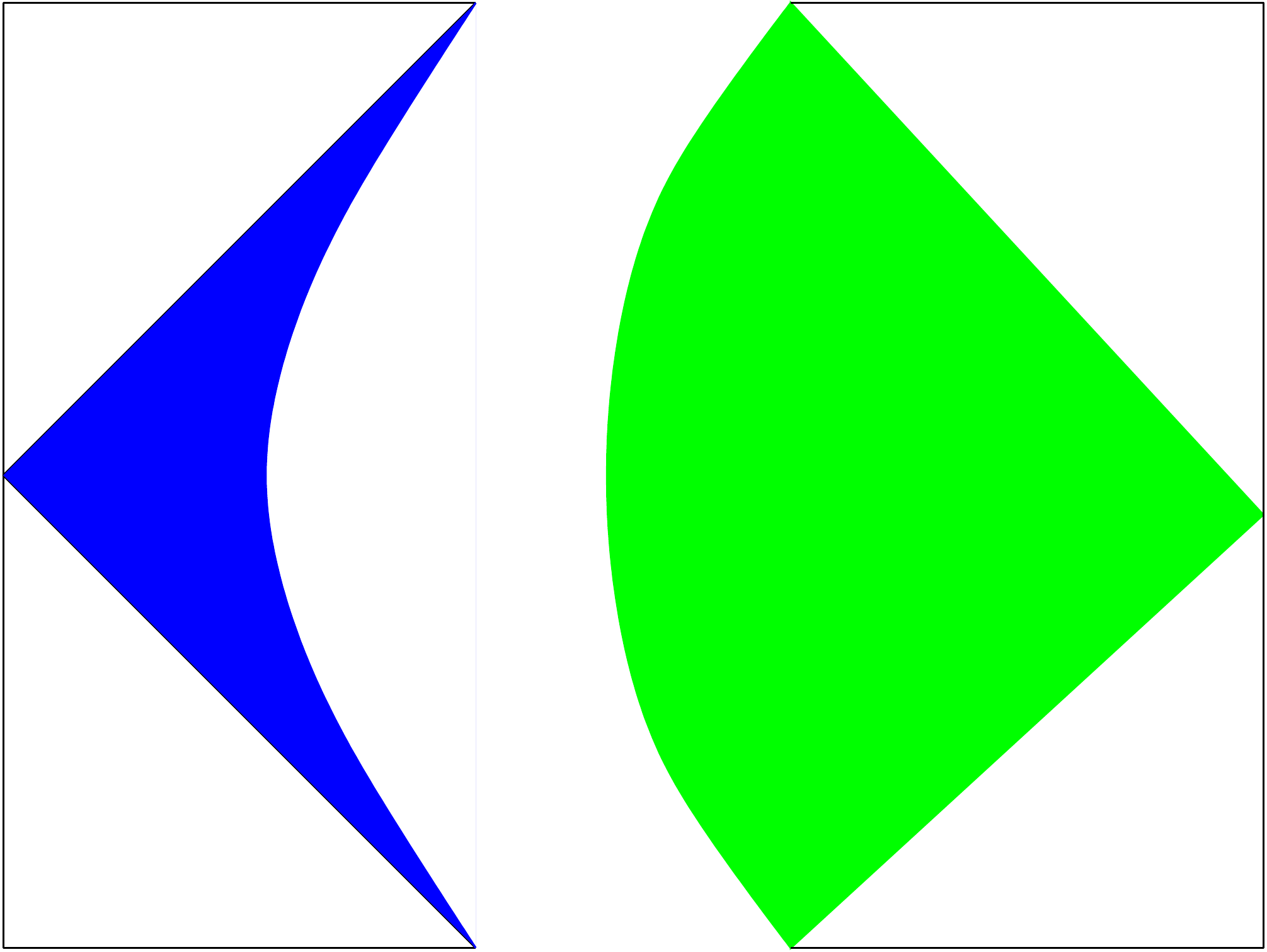}
\end{center}
\caption{The corresponding bounce geometry.  The charged brane sits on the
matching slice of two portions.  In the left portion the field $\phi$
follows the blue potential, and in the right portion it follows the green 
potential.  Our metric describes the shaded parts and can be analytically 
continued to obtain the rest of the geometry.}
 \label{fig-bounce}
\end{figure}

In Fig.\ref{fig-pot} we sketch the shape of effective potentials with 
different values of $Q$ and indeed found a local minimum as a stable 4D 
vacuum.  Unfortunately, different 4D vacua are present in different effective 
potentials, but conventional way to describe a transition only applies to
two vacua in the same potential.

However, the 6D description provides a simple answer.  As depicted in 
Fig.~\ref{fig-4t4}, \ref{fig-bounce}, the field $\phi$ follows the potential with $Q_1$ from its vacuum value until the position of the charged brane, 
$\bar{\phi}=2m_4\ln(m_6\bar{B})$, then jumps to the other potential with 
$Q_2$ and proceeds to the other vacuum.  

The velocity $\phi'$ will be discontinuous at the jump, but can be computed from the extra junction condition in 6D, Eq.~(\ref{eq-juncq}).
\begin{equation}
\phi_1'+\phi_2'=\frac{3\sigma}{\sqrt{4\pi}m_6^4\bar{B}^3}~.
\label{eq-juncphi}
\end{equation}

Since ideal 4D vacua are very close together, there will be a range where the 
following expansion holds in both potentials.
\begin{equation}
V_i(\phi)=\frac{12\pi H_i^2}{R_i^2}+
\frac{2\Lambda_6^2}{m_6^{10}}(\phi-\phi_i)^2~.
\end{equation}
Here $e^{\phi_i/2m_4}=m_6R_i$ is the vacuum value of the field.

\begin{figure}
\begin{center}
\includegraphics[scale = 0.35, angle=0]{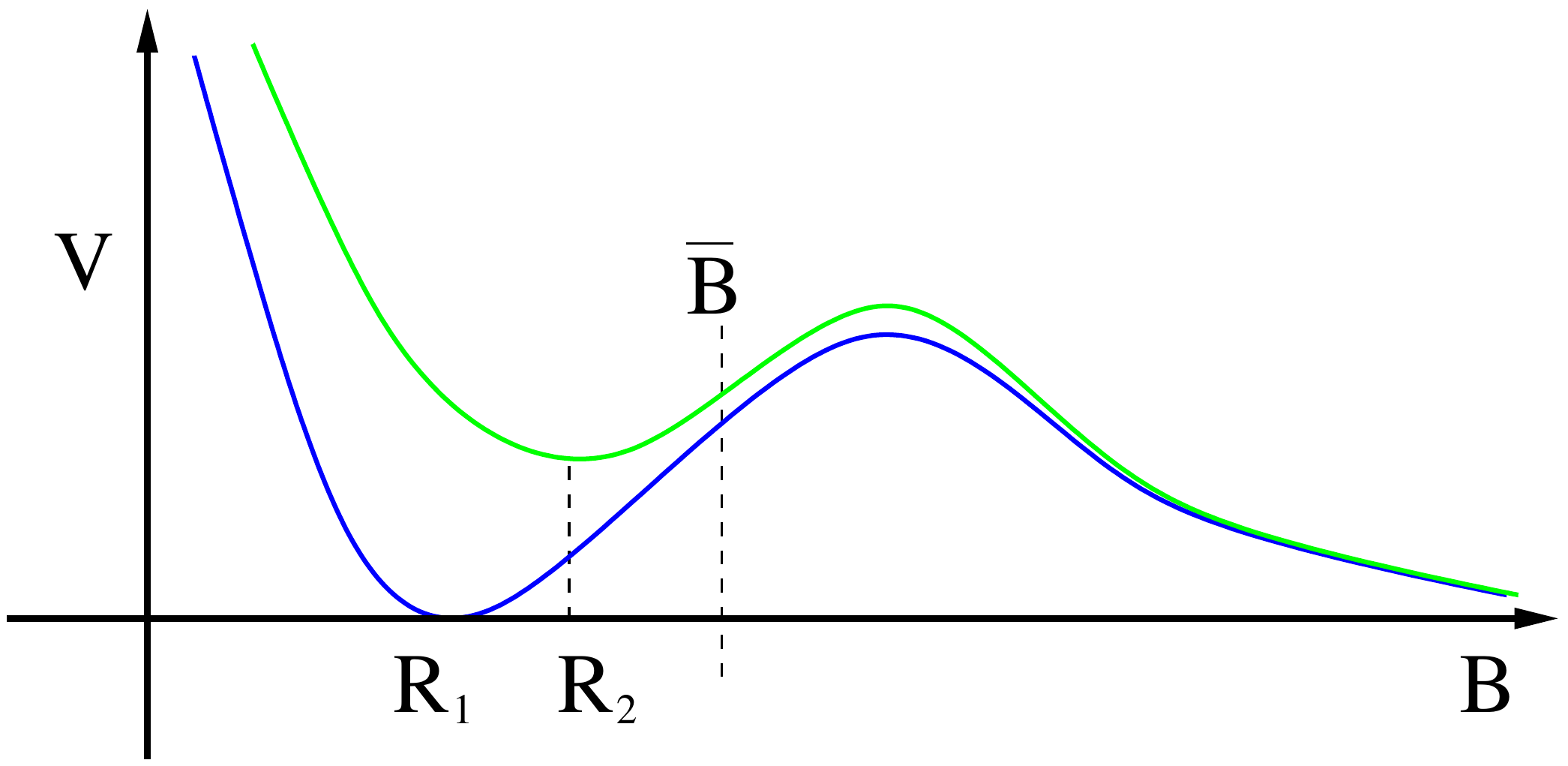}
\end{center}
\caption{The matching radius $\bar{B}$ can be larger than either vacuum 
values $R_i$. }
 \label{fig-stretched}
\begin{center}
\includegraphics[scale = 0.4, angle=0]{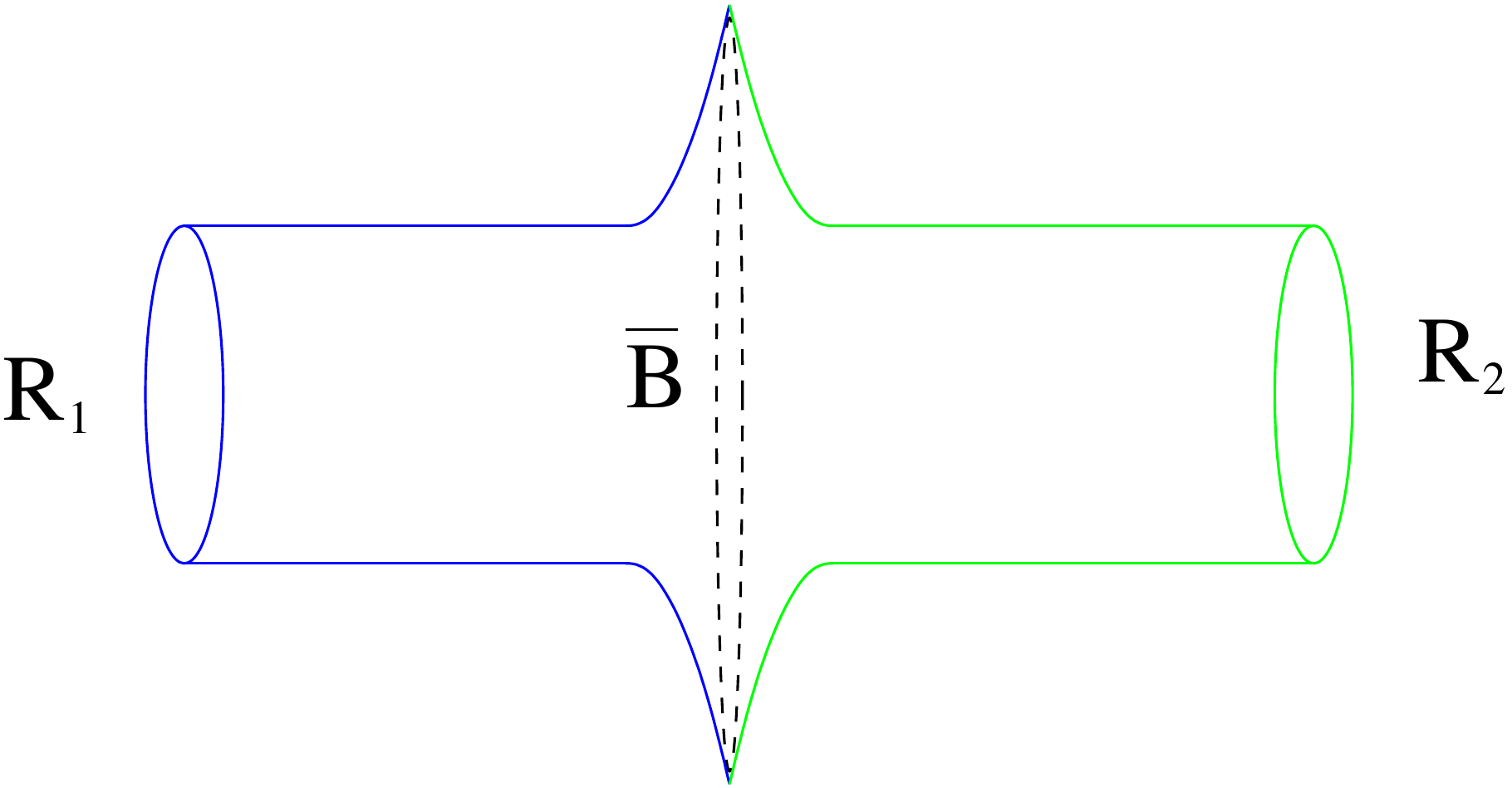}
\end{center}
\caption{The size of extra dimension $S_2$ gets stretched during the transition from vacuum 2(green, right) to vacuum 1(blue, left).}
 \label{fig-tube}
\end{figure}

We can see the mass of the potential is set by the 6D vacuum energy, which is 
much larger than the 4D vacuum energy that controls the bounce geometry.  By 
the argument of Coleman, we can ignore the friction term in 
Eq.~(\ref{eq-field}) and pretend the energy is conserved.  Together with the 
convention we mentioned earlier, that $\phi'$ is positive if it is increasing 
toward the brane, we have
\begin{equation}
\phi_i'=\frac{2\Lambda_6}{m_6^5}(\bar{\phi}-\phi_i)~,
\label{eq-approx}
\end{equation}
where $e^{\bar{\phi}/2m_4}=m_6\bar{B}$ is the matching value of the field.  
Plugging into the junction condition Eq.~(\ref{eq-juncphi}), we have
\begin{equation}
\bar{\phi}-\frac{\phi_1+\phi_2}{2}
=\frac{3\sigma m_6}{4\sqrt{4\pi}\bar{B}^3\Lambda_6}~.
\end{equation}

It is more illuminating to take exponential of the above equation.
\begin{equation}
\frac{\bar{B}^2}{R_1R_2}=
e^{\frac{3\sigma m_6}{4\sqrt{4\pi}m_4\bar{B}^3\Lambda_6}}~.
\label{eq-detour}
\end{equation}
The matching radius $\bar{B}$ has to be larger than the geometric mean of the
vacua radii for the tension to be positive.  Also, depending on the tension,
the transition can be monotonic as depicted in Fig.\ref{fig-4t4}, or with $\bar{B}>R_2$, in which the field goes to a larger value then comes back, 
as in Fig.\ref{fig-stretched}, the extra dimensions are stretched.  The 
critical tension when $\bar{B}=R_2$ is related to the charge $\Delta Q$ by 
Eq.~(\ref{eq-R}),(\ref{eq-detour}) and (\ref{eq-Q}).
\begin{equation}
\sigma_c=2\sqrt{2\pi}m_6^2\Delta Q~.
\label{eq-BPS}
\end{equation}

\section{4D to 6D solution}
\label{sec-4t6}

\begin{figure}[hb]
\begin{center}
\includegraphics[scale = 0.45, angle=0]{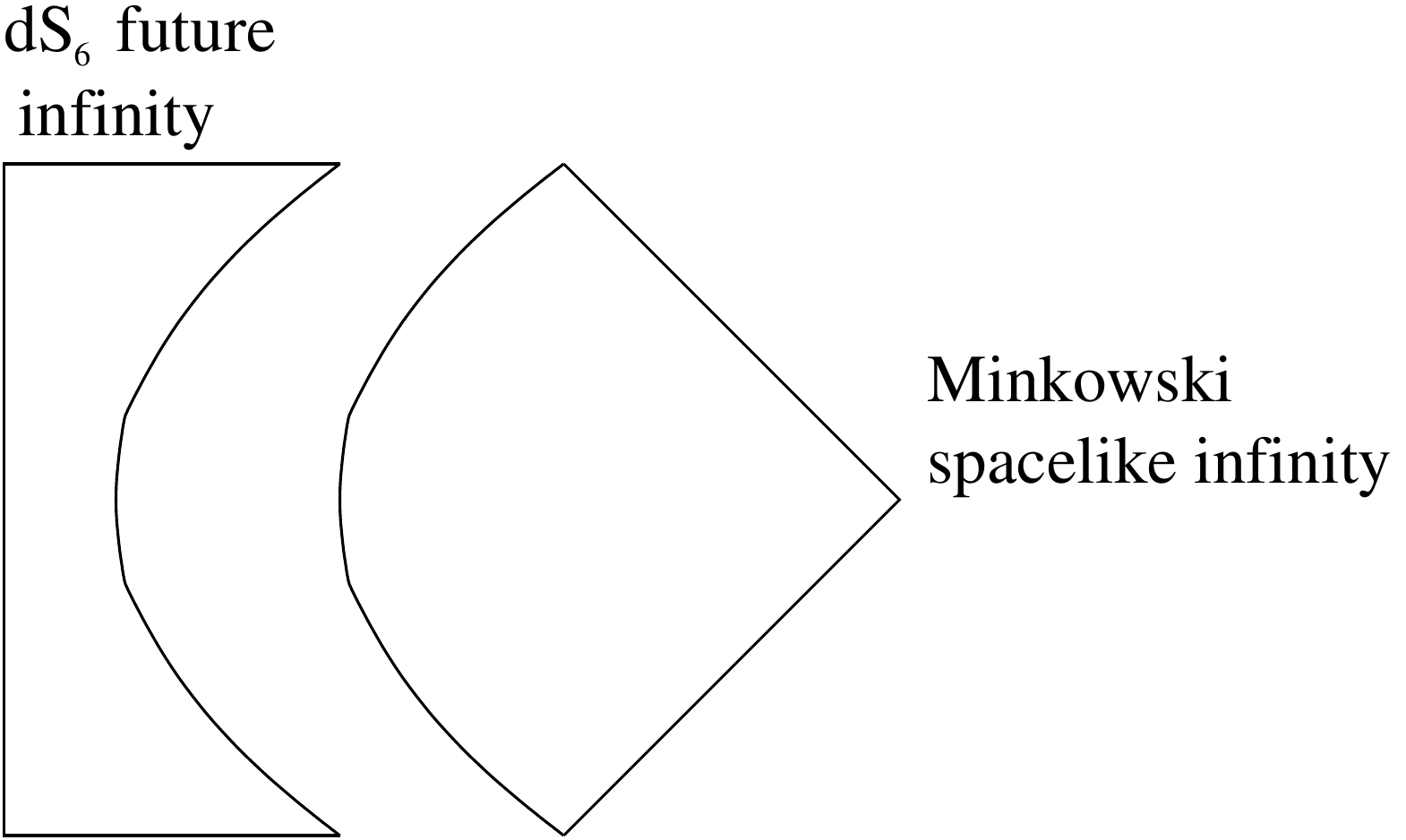}
\end{center}
\caption{Matching the interior of a small bubble in 4D Minkowski space to the 
bigger portion of $dS_6$(left hand side of Fig.\protect\ref{fig-6D}).  
Coexistence of Minkowski spacelike infinity and $dS_6$ future infinity 
violates the null energy condition.}
 \label{fig-big}
\begin{center}
\includegraphics[scale = 0.45, angle=0]{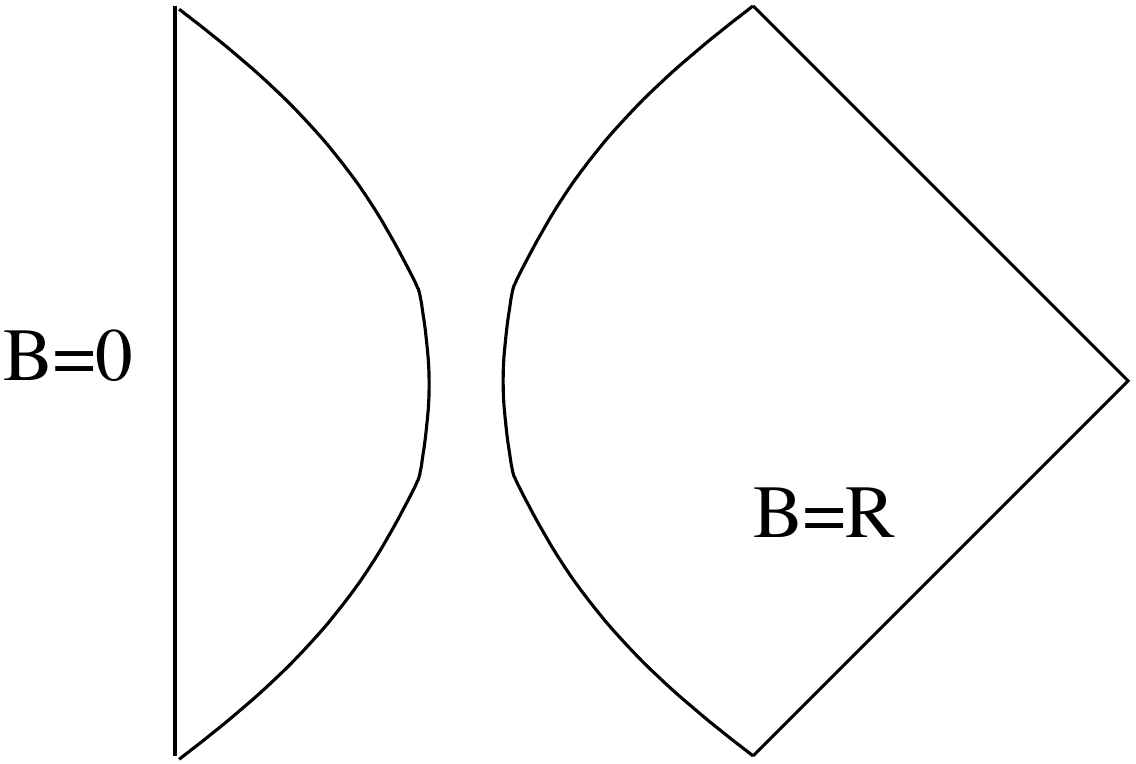}
\end{center}
\caption{Matching the interior of a small bubble in 4D Minkowski space to the 
smaller portion of $dS_6$(right hand side of Fig.\protect\ref{fig-6D}).  The 
extra dimension $S_2$ pinches off smoothly at the left end of this diagram.}
 \label{fig-small}
\end{figure}

Since the extra dimensions can be stretched during a transition, more 
dramatic effects like decompactifications might also happen.  We can use the 
same technique of jumping between potentials, but this time from a potential 
with $Q>0$ to $Q=0$.  Note that this is not covered in\cite{CarJoh09} 
where the entire solution is on one effective potential.  The most obvious 
difference is the charged brane appearing explicitly in our solution.
 
From Eq.~(\ref{eq-HQ}) we know that the 4D vacuum has to be outside, since it 
has the larger $Q$.  It means that the nontrivial region in 4D is only a 
small bubble, therefore there is no major difference between a deSitter, 
Minkowski or AdS space.  We shall take the 4D side to be Minkowski 
space for simplicity.  The 6D side has two possibilities.  Taking the left 
portion in Fig.\ref{fig-6D} we get Fig.\ref{fig-big}, while the right 
portion gives Fig.\ref{fig-small}.

Fig.\ref{fig-big} speaks trouble.  It contains a piece of deSitter future infinity surrounded by flat spacelike asymptotics.  By the arguement of 
Farhi and Guth in~\cite{FarGut87}, it needs to violate the null energy 
condition.  In Appendix~\ref{sec-case} we confirm this by showing explicitly
that the junction conditions lead to a domain wall with negative tension.

This leaves us with Fig.\ref{fig-small}, where the 6D piece is small and does 
not have deSitter asymptotics.  It is just a piece of spacetime letting the 
extradimensional $S_2$ pinch off in a non-singular way.  In Appendix
\ref{sec-case} we also show by junction conditions that such geometry really
exists.  Instead of a decompactification, we get a bubble of nothing
\cite{Wit81}\footnote{We thank Ben Freivogel for pointing this out.}!    

\section{Discussion}
\label{sec-dis}
 \subsection{Stretched extra dimensions}
 \label{sec-dis1}
 
In Sec.\ref{sec-4t4eff} we demonstrated a vacuum transition with extra
dimensions stretched, $(\bar{B}-R_i)\gg (R_2-R_1)$, in the sense that it
seems to get larger than necessary while interpolating between the vacuum 
values.  Within the validity range of our approximations, the sizes of the 
extra dimensions are still quite similar, $(\bar{B}-R_i)\ll R_i$.  It means 
that we can still understand the geometry in the 4D picture.  Also, the 
stretching process has the time scale of 6D physics, so from the 4D 
prospective it is very fast, therefore still a thin-wall solution.  We did 
not calculate the tunneling rate explicitly because it will not be 
significantly different from the pure 4D CDL\cite{CDL} tunneling.

If we can go beyond the approximation in Eq.~(\ref{eq-approx}), we can 
increase $\sigma$ even further and describe vacuum transitions with 
$\bar{B}\gg R_i$.  In which case it will be interesting to think about the tunneling rates and many other things.  In particular, there might be 
$\sigma$ too large that a transition is forbidden as conjectured 
in\cite{JohLar08}.

\begin{figure}
\begin{center}
\includegraphics[scale = 0.4, angle=0]{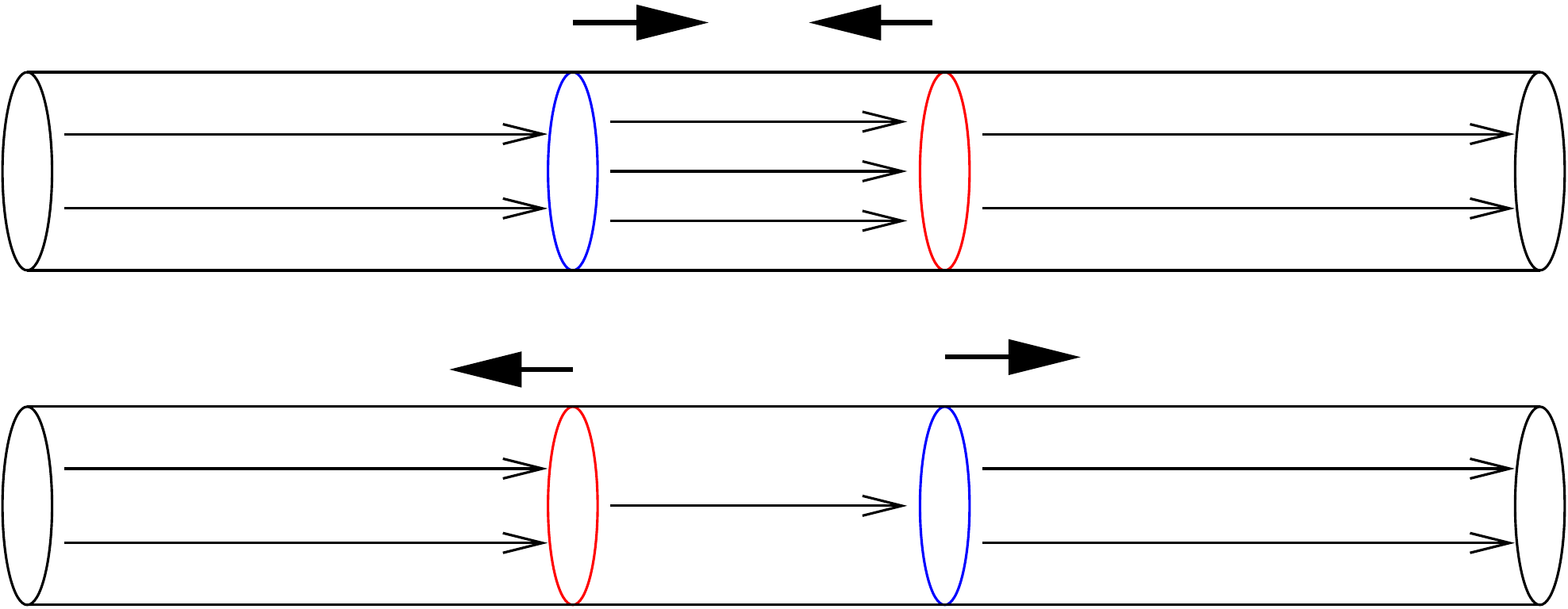}
\end{center}
\caption{Bubble collision when the extra dimensions unstretched.  The 
charged branes pass through each other and change the flux in the middle 
region, generating a third (lower) vacuum.  With extra dimensions stretched, 
the charged branes will not pass through each other this trivially.}
 \label{fig-col}
\end{figure}

A small stretch described here can already affect one thing---the 
bubble collisions.  An interesting conjecture in\cite{BlaJos09} says that 
the domain walls will pass through each other and leave a third vacuum in 
between, as shown in Fig.\ref{fig-col}.  It is supported by simulations of 
single-field-tunnelings\cite{EasGib09}.  The stretches we discovered in this 
paper unavoidably introduce an additional short-range interaction between 
domain walls and might change the conclusion.

The critical tension to stretch the extra dimensions in Eq.~(\ref{eq-BPS}) 
looks very similar to the familiar BPS bound in supersymmetric theories.  One 
may want to argue that for the real string theory landscape, since we started 
from a supersymmetric theory, the generic object will be much heavier than 
the bound, therefore always stretches the extra dimensions.

On the other hand, this toy model has 6D vacuum energy to start with.  One 
can also argue that gravity has to be the weakest force\cite{ArkMot06}
\footnote{We thank Robert Myers for pointing this out.}.  And there must be
a charged brane that does not stretch the extra dimensions.

We will not argue in favor of either side, but simply present this model as 
a tool to analyze the situation when the extra dimensions are stretched.
For the reasons given in the introduction, this might be an unavoidable 
situation.

 \subsection{Bubbles of nothing}
 \label{sec-dis2}

In Sec.\ref{sec-4t6} and Appendix~\ref{sec-case} we provided two pieces of 
evidence that a bubble of nothing appears in the place of a 
decompactification.  The Farhi-Guth\cite{FarGut87} arguement applies to the 
specific case in Fig.\ref{fig-big}, where inside the bubble is a deSitter 
space and outside is flat.  We are quite confident that the conclusion still holds when the exterior becomes $dS_4$ or $AdS_4$, because a small bubble 
effect does not care about asymptotics.

We should think about another case where we cannot apply the Farhi-Guth 
arguement---when the higher dimensional geometry is not deSitter, but flat as
in the string theory.  It will be very interesting if the junction conditions 
still work in a similar way as in Appendix~\ref{sec-case}, showing that 
positive tension demands a bubble of nothing instead of a decompactification.

We need to further study the decay rates.  But from the fact that we can arrange our solution to be a small bubble, it should have a rate very similar
to a standard thin-wall quantum instanton\cite{CDL,BroWei07,BlaJos09}.  Other 
geometries representing decompactification tunnelings all have rates close to 
a thermal instanton\cite{KKLT,GidMye04,CarJoh09,BlaJos09}.  This may suggest
that bubbles of nothing is the real universal instability we should think 
about, not decompactifications.

\acknowledgments 
We thank Ben Freivogel for numerous discussions and interesting suggestions 
that helped to shape this paper.  We also thank Raphael Bousso, Matthew 
Johnson and Robert Myers for stimulating discussions.  This work is supported 
in part by the US Department of Energy, and partially done in the Berkeley 
Center of Theoretical Physics.

%Andrei Linde and Alexander Westphal

\bibliographystyle{apsrev4-1}
\bibliography{all}

%Merlin.mbs v4.21 2009-07-09.
\begin{thebibliography}{10}%
\makeatletter
\providecommand \@ifxundefined [1]{%
 \ifx #1\undefined \expandafter \@firstoftwo
 \else \expandafter \@secondoftwo
\fi
}%
\providecommand \@ifnum [1]{%
 \ifnum #1\expandafter \@firstoftwo
 \else \expandafter \@secondoftwo
\fi
}%
\providecommand \enquote [1]{``#1''}%
\providecommand \bibnamefont  [1]{#1}%
\providecommand \bibfnamefont [1]{#1}%
\providecommand \citenamefont [1]{#1}%
\providecommand\href[0]{\@sanitize\@href}%
\providecommand\@href[1]{\endgroup\@@startlink{#1}\endgroup\@@href}%
\providecommand\@@href[1]{#1\@@endlink}%
\providecommand \@sanitize [0]{\begingroup\catcode`\&12\catcode`\#12\relax}%
\@ifxundefined \pdfoutput {\@firstoftwo}{%
 \@ifnum{\z@=\pdfoutput}{\@firstoftwo}{\@secondoftwo}%
}{%
 \providecommand\@@startlink[1]{\leavevmode\special{html:<a href="#1">}}%
 \providecommand\@@endlink[0]{\special{html:</a>}}%
}{%
 \providecommand\@@startlink[1]{%
  \leavevmode
  \pdfstartlink
   attr{/Border[0 0 1 ]/H/I/C[0 1 1]}%
   user{/Subtype/Link/A<</Type/Action/S/URI/URI(#1)>>}%
  \relax
 }%
 \providecommand\@@endlink[0]{\pdfendlink}%
}%
\providecommand \url  [0]{\begingroup\@sanitize \@url }%
\providecommand \@url [1]{\endgroup\@href {#1}{\urlprefix}}%
\providecommand \urlprefix [0]{URL }%
\providecommand \Eprint[0]{\href }%
\@ifxundefined \urlstyle {%
  \providecommand \doi [1]{doi:\discretionary{}{}{}#1}%
}{%
  \providecommand \doi [0]{doi:\discretionary{}{}{}\begingroup
  \urlstyle{rm}\Url }%
}%
\providecommand \doibase [0]{http://dx.doi.org/}%
\providecommand \Doi[1]{\href{\doibase#1}}%
\providecommand \bibAnnote [3]{%
  \BibitemShut{#1}%
  \begin{quotation}\noindent
    \textsc{Key:}\ #2\\\textsc{Annotation:}\ #3%
  \end{quotation}%
}%
\providecommand \bibAnnoteFile [2]{%
  \IfFileExists{#2}{\bibAnnote {#1} {#2} {\input{#2}}}{}%
}%
\providecommand \typeout [0]{\immediate \write \m@ne }%
\providecommand \selectlanguage [0]{\@gobble}%
\providecommand \bibinfo [0]{\@secondoftwo}%
\providecommand \bibfield [0]{\@secondoftwo}%
\providecommand \translation [1]{[#1]}%
\providecommand \BibitemOpen[0]{}%
\providecommand \bibitemStop [0]{}%
\providecommand \bibitemNoStop [0]{.\EOS\space}%
\providecommand \EOS [0]{\spacefactor3000\relax}%
\providecommand \BibitemShut [1]{\csname bibitem#1\endcsname}%
%</preamble>
\bibitem{FreRub80}%
  \BibitemOpen
  \bibfield{author}{%
  \bibinfo {author} {\bibfnamefont{P.~G.~O.}\ \bibnamefont{Freund}}\ and\
  \bibinfo {author} {\bibfnamefont{M.~A.}\ \bibnamefont{Rubin}},\ }%
  \bibfield{journal}{%
  \Doi{10.1016/0370-2693(80)90590-0}{\bibinfo {journal} {Phys. Lett.}}\ }%
  \textbf{\bibinfo {volume} {B97}},\ \bibinfo {pages} {233} (\bibinfo {year}
  {1980})%
  \bibAnnoteFile{NoStop}{FreRub80}%
%%CITATION = PHLTA,B97,233;%%
\bibitem{RanSal82}%
  \BibitemOpen
  \bibfield{author}{%
  \bibinfo {author} {\bibfnamefont{S.}~\bibnamefont{Randjbar-Daemi}}, \bibinfo
  {author} {\bibfnamefont{A.}~\bibnamefont{Salam}},\ and\ \bibinfo {author}
  {\bibfnamefont{J.~A.}\ \bibnamefont{Strathdee}},\ }%
  \bibfield{journal}{%
  \Doi{10.1016/0550-3213(83)90247-X}{\bibinfo {journal} {Nucl. Phys.}}\ }%
  \textbf{\bibinfo {volume} {B214}},\ \bibinfo {pages} {491} (\bibinfo {year}
  {1983})%
  \bibAnnoteFile{NoStop}{RanSal82}%
%%CITATION = NUPHA,B214,491;%%
\bibitem{DouKac06}%
  \BibitemOpen
  \bibfield{author}{%
  \bibinfo {author} {\bibfnamefont{M.~R.}\ \bibnamefont{Douglas}}\ and\
  \bibinfo {author} {\bibfnamefont{S.}~\bibnamefont{Kachru}}}%
   (\bibinfo {year} {2006}),\
  \Eprint{http://arxiv.org/abs/hep-th/0610102}{hep-th/0610102}%
  \bibAnnoteFile{NoStop}{DouKac06}%
%%CITATION = HEP-TH 0610102;%%
\bibitem{BlaJos09}%
  \BibitemOpen
  \bibfield{author}{%
  \bibinfo {author} {\bibfnamefont{J.~J.}\ \bibnamefont{Blanco-Pillado}},
  \bibinfo {author} {\bibfnamefont{D.}~\bibnamefont{Schwartz-Perlov}},\ and\
  \bibinfo {author} {\bibfnamefont{A.}~\bibnamefont{Vilenkin}}}%
   (\bibinfo {year} {2009}),\
  \Eprint{http://arxiv.org/abs/0904.3106}{arXiv:0904.3106 [hep-th]}%
  \bibAnnoteFile{NoStop}{BlaJos09}%
%%CITATION = 0904.3106;%%
\bibitem{CarJoh09}%
  \BibitemOpen
  \bibfield{author}{%
  \bibinfo {author} {\bibfnamefont{S.~M.}\ \bibnamefont{Carroll}}, \bibinfo
  {author} {\bibfnamefont{M.~C.}\ \bibnamefont{Johnson}},\ and\ \bibinfo
  {author} {\bibfnamefont{L.}~\bibnamefont{Randall}}}%
   (\bibinfo {year} {2009}),\
  \Eprint{http://arxiv.org/abs/0904.3115}{arXiv:0904.3115 [hep-th]}%
  \bibAnnoteFile{NoStop}{CarJoh09}%
%%CITATION = 0904.3115;%%
\bibitem{BP}%
  \BibitemOpen
  \bibfield{author}{%
  \bibinfo {author} {\bibfnamefont{R.}~\bibnamefont{Bousso}}\ and\ \bibinfo
  {author} {\bibfnamefont{J.}~\bibnamefont{Polchinski}},\ }%
  \bibfield{journal}{%
  \bibinfo {journal} {JHEP}\ }%
  \textbf{\bibinfo {volume} {06}},\ \bibinfo {pages} {006} (\bibinfo {year}
  {2000}),\ \Eprint{http://arxiv.org/abs/hep-th/0004134}{hep-th/0004134}%
  \bibAnnoteFile{NoStop}{BP}%
%%CITATION = JHEPA,0006,006;%%
\bibitem{JohLar08}%
  \BibitemOpen
  \bibfield{author}{%
  \bibinfo {author} {\bibfnamefont{M.~C.}\ \bibnamefont{Johnson}}\ and\
  \bibinfo {author} {\bibfnamefont{M.}~\bibnamefont{Larfors}},\ }%
  \bibfield{journal}{%
  \Doi{10.1103/PhysRevD.78.123513}{\bibinfo {journal} {Phys. Rev.}}\ }%
  \textbf{\bibinfo {volume} {D78}},\ \bibinfo {pages} {123513} (\bibinfo {year}
  {2008}),\ \Eprint{http://arxiv.org/abs/0809.2604}{arXiv:0809.2604 [hep-th]}%
  \bibAnnoteFile{NoStop}{JohLar08}%
%%CITATION = 0809.2604;%%
\bibitem{GidMye04}%
  \BibitemOpen
  \bibfield{author}{%
  \bibinfo {author} {\bibfnamefont{S.~B.}\ \bibnamefont{Giddings}}\ and\
  \bibinfo {author} {\bibfnamefont{R.~C.}\ \bibnamefont{Myers}},\ }%
  \bibfield{journal}{%
  \Doi{10.1103/PhysRevD.70.046005}{\bibinfo {journal} {Phys. Rev.}}\ }%
  \textbf{\bibinfo {volume} {D70}},\ \bibinfo {pages} {046005} (\bibinfo {year}
  {2004}),\ \Eprint{http://arxiv.org/abs/hep-th/0404220}{arXiv:hep-th/0404220}%
  \bibAnnoteFile{NoStop}{GidMye04}%
%%CITATION = HEP-TH/0404220;%%
\bibitem{Wit81}%
  \BibitemOpen
  \bibfield{author}{%
  \bibinfo {author} {\bibfnamefont{E.}~\bibnamefont{Witten}},\ }%
  \bibfield{journal}{%
  \Doi{10.1016/0550-3213(82)90007-4}{\bibinfo {journal} {Nucl. Phys.}}\ }%
  \textbf{\bibinfo {volume} {B195}},\ \bibinfo {pages} {481} (\bibinfo {year}
  {1982})%
  \bibAnnoteFile{NoStop}{Wit81}%
%%CITATION = NUPHA,B195,481;%%
\bibitem{EasGib09}%
  \BibitemOpen
  \bibfield{author}{%
  \bibinfo {author} {\bibfnamefont{R.}~\bibnamefont{Easther}}, \bibinfo
  {author} {\bibfnamefont{J.}~\bibnamefont{Giblin}, \bibfnamefont{John~T.}},
  \bibinfo {author} {\bibfnamefont{L.}~\bibnamefont{Hui}},\ and\ \bibinfo
  {author} {\bibfnamefont{E.~A.}\ \bibnamefont{Lim}}}%
   (\bibinfo {year} {2009}),\
  \Eprint{http://arxiv.org/abs/0907.3234}{arXiv:0907.3234 [hep-th]}%
  \bibAnnoteFile{NoStop}{EasGib09}%
%%CITATION = 0907.3234;%%
\bibitem{KKLT}%
  \BibitemOpen
  \bibfield{author}{%
  \bibinfo {author} {\bibfnamefont{S.}~\bibnamefont{Kachru}}, \bibinfo {author}
  {\bibfnamefont{R.}~\bibnamefont{Kallosh}}, \bibinfo {author}
  {\bibfnamefont{A.}~\bibnamefont{Linde}},\ and\ \bibinfo {author}
  {\bibfnamefont{S.~P.}\ \bibnamefont{Trivedi}},\ }%
  \bibfield{journal}{%
  \bibinfo {journal} {Phys. Rev. D}\ }%
  \textbf{\bibinfo {volume} {68}},\ \bibinfo {pages} {046005} (\bibinfo {year}
  {2003}),\ \Eprint{http://arxiv.org/abs/hep-th/0301240}{hep-th/0301240}%
  \bibAnnoteFile{NoStop}{KKLT}%
%%CITATION = HEP-TH 0301240;%%
\bibitem{BlaBur04}%
  \BibitemOpen
  \bibfield{author}{%
  \bibinfo {author} {\bibfnamefont{J.~J.}\ \bibnamefont{Blanco-Pillado}}
  \emph{et~al.},\ }%
  \bibfield{journal}{%
  \Doi{10.1088/1126-6708/2004/11/063}{\bibinfo {journal} {JHEP}}\ }%
  \textbf{\bibinfo {volume} {11}},\ \bibinfo {pages} {063} (\bibinfo {year}
  {2004}),\ \Eprint{http://arxiv.org/abs/hep-th/0406230}{arXiv:hep-th/0406230}%
  \bibAnnoteFile{NoStop}{BlaBur04}%
%%CITATION = HEP-TH/0406230;%%
\bibitem{BouDew02}%
  \BibitemOpen
  \bibfield{author}{%
  \bibinfo {author} {\bibfnamefont{R.}~\bibnamefont{Bousso}}, \bibinfo {author}
  {\bibfnamefont{O.}~\bibnamefont{DeWolfe}},\ and\ \bibinfo {author}
  {\bibfnamefont{R.~C.}\ \bibnamefont{Myers}},\ }%
  \bibfield{journal}{%
  \bibinfo {journal} {Found. Phys.}\ }%
  \textbf{\bibinfo {volume} {33}},\ \bibinfo {pages} {297} (\bibinfo {year}
  {2003}),\ \Eprint{http://arxiv.org/abs/hep-th/0205080}{hep-th/0205080}%
  \bibAnnoteFile{NoStop}{BouDew02}%
%%CITATION = HEP-TH 0205080;%%
\bibitem{KriPab05}%
  \BibitemOpen
  \bibfield{author}{%
  \bibinfo {author} {\bibfnamefont{C.}~\bibnamefont{Krishnan}}, \bibinfo
  {author} {\bibfnamefont{S.}~\bibnamefont{Paban}},\ and\ \bibinfo {author}
  {\bibfnamefont{M.}~\bibnamefont{Zanic}},\ }%
  \bibfield{journal}{%
  \bibinfo {journal} {JHEP}\ }%
  \textbf{\bibinfo {volume} {05}},\ \bibinfo {pages} {045} (\bibinfo {year}
  {2005}),\ \Eprint{http://arxiv.org/abs/hep-th/0503025}{arXiv:hep-th/0503025}%
  \bibAnnoteFile{NoStop}{KriPab05}%
%%CITATION = HEP-TH/0503025;%%
\bibitem{Isr66}%
  \BibitemOpen
  \bibfield{author}{%
  \bibinfo {author} {\bibfnamefont{W.}~\bibnamefont{Israel}},\ }%
  \bibfield{journal}{%
  \Doi{10.1007/BF02730328}{\bibinfo {journal} {Nuovo Cim.}}\ }%
  \textbf{\bibinfo {volume} {B44S10}},\ \bibinfo {pages} {1} (\bibinfo {year}
  {1966})%
  \bibAnnoteFile{NoStop}{Isr66}%
%%CITATION = NUCIA,B44S10,1;%%
\bibitem{CDL}%
  \BibitemOpen
  \bibfield{author}{%
  \bibinfo {author} {\bibfnamefont{S.}~\bibnamefont{Coleman}}\ and\ \bibinfo
  {author} {\bibfnamefont{F.~D.}\ \bibnamefont{Luccia}},\ }%
  \bibfield{journal}{%
  \bibinfo {journal} {Phys. Rev. D}\ }%
  \textbf{\bibinfo {volume} {21}},\ \bibinfo {pages} {3305} (\bibinfo {year}
  {1980})%
  \bibAnnoteFile{NoStop}{CDL}%
\bibitem{FarGut87}%
  \BibitemOpen
  \bibfield{author}{%
  \bibinfo {author} {\bibfnamefont{E.}~\bibnamefont{Farhi}}\ and\ \bibinfo
  {author} {\bibfnamefont{A.~H.}\ \bibnamefont{Guth}},\ }%
  \bibfield{journal}{%
  \bibinfo {journal} {Phys. Lett.}\ }%
  \textbf{\bibinfo {volume} {B183}},\ \bibinfo {pages} {149} (\bibinfo {year}
  {1987})%
  \bibAnnoteFile{NoStop}{FarGut87}%
%%CITATION = PHLTA,B183,149;%%
\bibitem{ArkMot06}%
  \BibitemOpen
  \bibfield{author}{%
  \bibinfo {author} {\bibfnamefont{N.}~\bibnamefont{Arkani-Hamed}}, \bibinfo
  {author} {\bibfnamefont{L.}~\bibnamefont{Motl}}, \bibinfo {author}
  {\bibfnamefont{A.}~\bibnamefont{Nicolis}},\ and\ \bibinfo {author}
  {\bibfnamefont{C.}~\bibnamefont{Vafa}},\ }%
  \bibfield{journal}{%
  \bibinfo {journal} {JHEP}\ }%
  \textbf{\bibinfo {volume} {06}},\ \bibinfo {pages} {060} (\bibinfo {year}
  {2007}),\ \Eprint{http://arxiv.org/abs/hep-th/0601001}{hep-th/0601001}%
  \bibAnnoteFile{NoStop}{ArkMot06}%
%%CITATION = HEP-TH/0601001;%%
\bibitem{BroWei07}%
  \BibitemOpen
  \bibfield{author}{%
  \bibinfo {author} {\bibfnamefont{A.~R.}\ \bibnamefont{Brown}}\ and\ \bibinfo
  {author} {\bibfnamefont{E.~J.}\ \bibnamefont{Weinberg}},\ }%
  \bibfield{journal}{%
  \Doi{10.1103/PhysRevD.76.064003}{\bibinfo {journal} {Phys. Rev.}}\ }%
  \textbf{\bibinfo {volume} {D76}},\ \bibinfo {pages} {064003} (\bibinfo {year}
  {2007}),\ \Eprint{http://arxiv.org/abs/0706.1573}{arXiv:0706.1573 [hep-th]}%
  \bibAnnoteFile{NoStop}{BroWei07}%
%%CITATION = 0706.1573;%%
\end{thebibliography}%

\appendix

\section{A case study for 4D to 6D transitions}
\label{sec-case}

In Sec.\ref{sec-4t4} we followed a general process to find the matching 
geometry with given $Q_1,Q_2,\Lambda_6$ and $\sigma$, which relies heavily on the approximation in Eq.~(\ref{eq-approx}), and the fact that on either side 
we can find a family of well-behaved solutions all similar to the 4D 
stablized geometry.

Unfortunately, Eq.~(\ref{eq-approx}) does not hold anywhere on the $Q=0$ 
potential, and even the na\"ive $dS_6$ solution in Eq.~(\ref{eq-dS6}) looks 
dangerously singular in the 4D equations of motion~(\ref{eq-field}).  A very 
careful numerical study might solve the problem, but we would like to do 
something slightly different here.

Instead of searching for geometries with all parameters fixed, we would like 
to specify the geometry at the $Q=0$ side.  As a trade off, the tension 
$\sigma$ cannot be fixed anymore.  We will show that in order to get the 
geometry in Fig.\ref{fig-big}, the tension has to be negative.  This agrees 
with our reasoning in Sec.\ref{sec-4t6} that it violates the null energy 
condition.  For the geometry in Fig.\ref{fig-small}, the solution works with 
a positive tension.

The $Q=0$ geometry will be specified as the pure $dS_6$ solution in 
Eq.~\ref{eq-dS6}.  With the convention of keeping the small $\rho$ region, 
it is like Fig.\ref{fig-small}.  
\begin{eqnarray}
A_1(\rho)&=&L\cos\frac{\rho}{L}~, 
\nonumber \\
B_1(\rho)&=&L\sin\frac{\rho}{L}~. 
\label{eq-small}
\end{eqnarray}
For Fig.\ref{fig-big}, we only need to exchange $A$ and $B$.
\begin{eqnarray}
A_1(\rho)&=&L\sin\frac{\rho}{L}~,
\nonumber \\
B_1(\rho)&=&L\cos\frac{\rho}{L}~.
\label{eq-big}
\end{eqnarray}

From the junction conditions, Eq.~(\ref{eq-juncq}), we have
\begin{eqnarray}
\dot{A}_2 &=& -\frac{\bar{A}\sigma}{16\pi\bar{B}^2m_6^4}-\dot{A}_1~, 
\label{eq-A2} \\
\dot{B}_2 &=& \frac{3\bar{B}\sigma}{16\pi\bar{B}^2m_6^4}-\dot{B}_1~.
\label{eq-B2}
\end{eqnarray}

By Eq.~(\ref{eq-Q}), $Q_2=m_6^4/\sqrt{\Lambda_6}$ has the 4D Minkowski vacuum.
The effective potential is
\begin{equation}
V(\bar{B})=\frac{2\pi}{R^2\bar{B}^2}\bigg(1-\frac{R^2}{\bar{B}^2}\bigg)^2~.
\end{equation}
For this potential we can still use the same approximation in 
Eq.~(\ref{eq-approx}).
\begin{equation}
\phi'=\frac{2\sqrt{\pi}}{R\bar{B}}\bigg(1-\frac{R^2}{\bar{B}^2}\bigg)~.
\end{equation}
Relate $\phi'$ to $\dot{B}_2$ and use Eq.~(\ref{eq-B2}), we have
\begin{equation}
\dot{B}_1+\frac{\bar{B}}{2R}\bigg(1-\frac{R^2}{\bar{B}^2}\bigg)
=\frac{3\sigma}{16\pi\bar{B}m_6^4}~.
\label{eq-1}
\end{equation}

As a small bubble in 4D Minkowski space, 
\begin{equation}
\bar{A}\frac{\dot{B}_2}{\bar{B}}+\dot{A}_2=a'=-1~.  
\end{equation}
Combine with Eq.~(\ref{eq-A2}) and (\ref{eq-B2}), we have 
\begin{equation}
\frac{3\bar{B}}{2\bar{A}}
\bigg(\dot{A_1}+\frac{\bar{A}}{\bar{B}}\dot{B_1}-1 \bigg)
=\frac{3\sigma}{16\pi\bar{B}m_6^4}~. 
\label{eq-2}
\end{equation}

Note that Eq.~(\ref{eq-1}) and Eq.~(\ref{eq-2}) have the same right hand side.
Also, since $L=\sqrt{10}m_6^2/\sqrt{\Lambda_6}=2\sqrt{5}R$, the lefthand sides
are just $\sin$ and $\cos$.  Plotting them between $0$ and $\pi/2$, we found 
that with Eq.~(\ref{eq-small}) they cross at a positive $\sigma$, but with 
Eq.~(\ref{eq-big}) they cross at a negative $\sigma$.  This confirms the 
arguement in Sec.\ref{sec-4t6} that Fig.\ref{fig-big} violates the null 
energy condition, and the bubble of nothing geometry in Fig.\ref{fig-small} 
really exists.

\begin{figure}
\begin{center}
\includegraphics[scale = 0.7, angle=0]{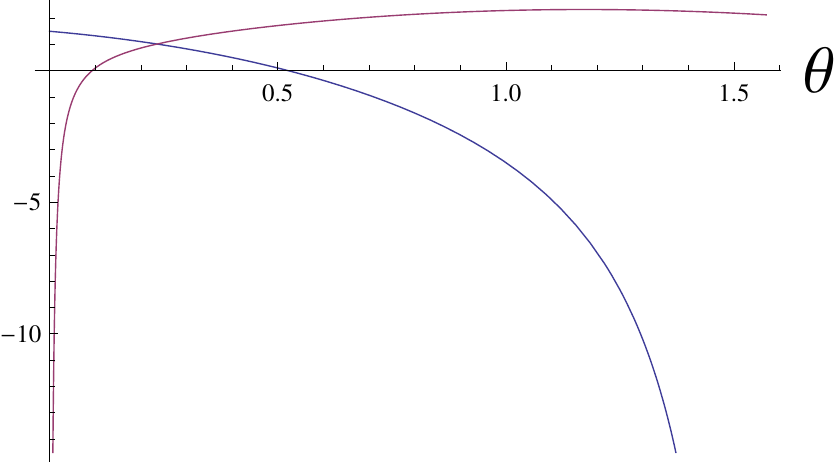}
\end{center}
\caption{The two junction conditions for the bubble of nothing geometry.  
They cross at a positive $\sigma$.}
 \label{fig-matchsmall}
\begin{center}
\includegraphics[scale = 0.7, angle=0]{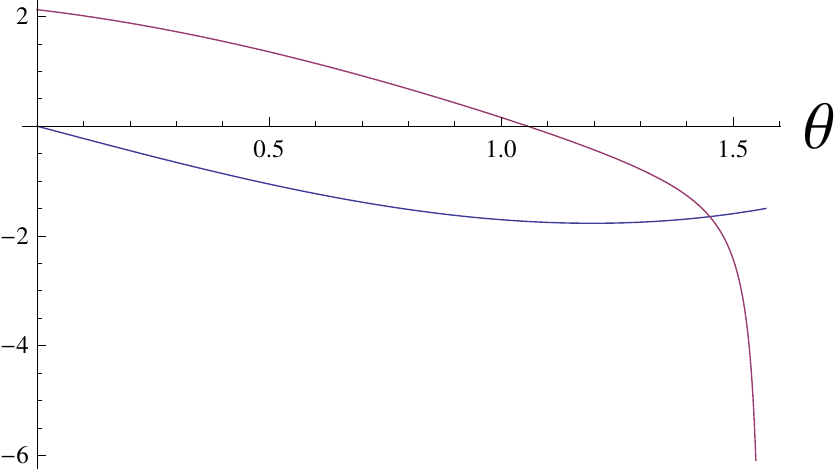}
\end{center}
\caption{The two junction conditions for the decompactification geometry.
They cross at a negative $\sigma$.}
 \label{fig-matchbig}
\end{figure}

\end{document}